\documentclass[review]{elsarticle}

\usepackage{hyperref}
\usepackage{xcolor}

\journal{Powder Technology}

\begin{document}

\begin{frontmatter}

\title{Flow of lubricated granular material on an inclined plane}

% \address{Radarweg 29, Amsterdam}
% \fntext[myfootnote]{Since 1880.}

\author[address1,address3]{Ravindra S. Ghodake}
\author[address2]{Pankaj Doshi\corref{}}
\cortext[]{Corresponding authors}
\ead{pankaj.doshi@pfizer.com}
\author[address1,address3]{Ashish V. Orpe\corref{}}
\ead{av.orpe@ncl.res.in}
\address[address1]{Chemical Engineering and Process Development
  Division, CSIR-National Chemical Laboratory, Pune 411008
India}
\address[address2]{Worldwide Research and Development, Pfizer Products
India Private Limited, Mumbai 400051 India}
\address[address3]{Academy of Scientific and Innovative Research
  (AcSIR), Ghaziabad 201002 India}

\begin{abstract}
  We have studied the gravity driven flow of spherical shaped,
  millimetric sized granular material coated with aspherical,
  micron-sized, near frictionless lubricant particles. Experiments
  were performed on an inclined plane using two different sized
  particles for varying concentrations of the lubricant. The particle
  volumetric flow rate exhibits a non-monotonic behavior with
  increasing lubricant concentration. It shows an increase at smaller
  lubricant concentration followed by a decrease at higher lubricant
  concentration. The lubricant particles adheres to the granular
  particle surface thereby reducing the inter-particle
  friction. However, presence of lubricant particles at higher
  concentration damps out inter-particle collision thereby reducing
  the inter-particle momentum transfer. Non-monotonicity in the
  observed behavior is then conjectured to arise due to competing
  effects of inter-particle friction and inter-particle collision. The
  present work and the overall observed behavior therein provides a
  simple experimental system to characterise the effects of added
  lubricant material in pharmaceutical and other relevant industrial
  applications.
\end{abstract}

\begin{keyword}
granular, powder, lubrication, friction, inclined plane
\end{keyword}

\end{frontmatter}

%\linenumbers

\section{\label{intro}Introduction}

Can we simply alter the inter-particle interaction, a fundamental
quantity governing the overall behavior in a dry granular system?
Doing so essentially will require either some kind of chemistry to
alter the particle surface or altering the material itself.  Both
these methods will change the inter-particle friction and collision
characteristics thereby modifying the flow. However, both the methods pose
restrictions in terms of material availability and limited scope and
ease in use of chemical treatment. Alternatively, DEM simulations
allow for studying dry granular systems by suitable tuning of
inter-particle friction and collisional
properties~\cite{silbert01,silbert03,orpe19}. But the technique still
presents limitations in terms of availability of experimental studies
for a detailed comparison in spite of several reported
studies~\cite{midi04}

As it turns out, addition of plate-like particles (typically Magnesium
Stearate or MgSt) alters the inter-particle interactions in dry
systems substantially~\cite{sun11}. This technique is used routinely
in pharmaceutical industries handling granular material in various
forms. The primary aim is to reduce the friction in the compression
die during the process of tablet compaction. As an additional benefit
the presence of MgSt also enhances the powder flowability.  Typically,
the MgSt powder is added in tiny quantities (weight ratio of
$O [10^{-2}]$). The individual particles, about $10$ micron sized,
adhere to the large particles and are expected to reduce the friction
between them due to their own frictionless nature thereby enhancing
the flowability. This flowability, using MgSt as lubricant, has been
studied previously to a reasonable
effect~\cite{llusa05,navaneethan05,kushner10,lakio13,morin13,wang17,ketterhagen18}. However,
all these studies have primarily focused on characterising the
optimality in terms of lubricant concentration vis-a-vis flow
characteristics. One of them have also focussed on studying the
avalanching and angle of repose behavior of pharmaceutical powders in
presence of various types of lubricants~\cite{morin13}. It was
observed that the addition of lubricant decreases the angle of repose
and the time for avalanching. The observed behavior was explained
based on particle morphology using scanning electron microscopy (SEM)
imaging.  On the other hand usage of powder as a lubricant between two
planar solid surfaces is very well known within the domain of
tribology and its merits and demerits with respect to conventional
liquid lubricants have been well documented~\cite{wornyoh07}. However,
the carryover of the flow~\cite{fillot05, iordanoff02} and
rheology~\cite{wornyoh05} characteristics from these planar surface
studies to bulk flow of lubricant-granular particles is not known.

Over here we attempt to connect the presence of lubricant particles,
their influence on the inter-particle interactions and eventually on
the shear flow behavior in terms of flow and concentration profiles
using flow visualization techniques. Given that MgSt particles are
soft in nature, their excessive presence and effectively higher
coating on the larger particles can render the latter to be soft,
thereby inducing a higher loss of momentum during collisions. This may
effectively nullify the advantage gained due to friction reducing
properties of the lubricant. We strive to understand these possible
contrasting behaviors in this work through detailed
experimentation. To obtain the necessary shear flow, we have chosen
chute (inclined plane) system, which is very well studied in
literature~\cite{midi04,pouliquen99,silbert01,jop06,silbert03,mills99,cassar05,campbell85}
covering various flow aspects and is, moreover, simple in its handling
and usage. The choice of the experimental system is purely incidental
and the purpose of this work is not to delve into the details of
inclined plane mechanics, but simply to create a shear flow which will
allow for studying and understanding the lubricant influence on the
flow behavior. 

\section{\label{method}Methodology}

\begin{figure}
  \centering
  \includegraphics[width=1.0\linewidth]{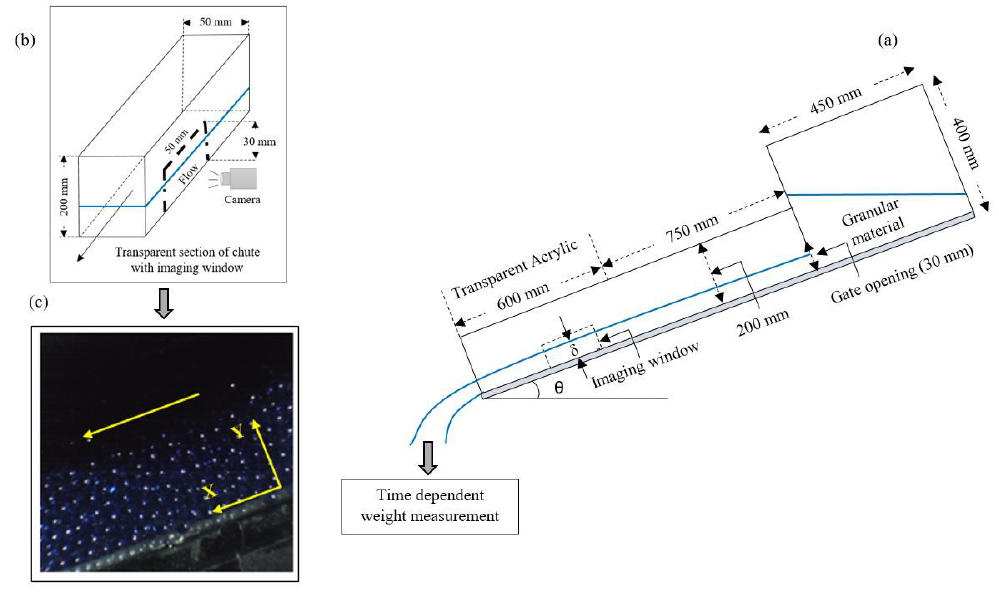}
  \caption{Schematic of the chute flow system. (a) Side view
    exhibiting hopper, chute, collection system and relevant
    dimensions. (b) Imaging of the flow at one of the chute side wall
    (c) Sample image captured using a high speed camera. See text
    for more details}
  \label{schem}
\end{figure}

Experiments are performed on a chute inclined at an
angle ($\theta$) as shown in fig.~\ref{schem}. The system comprises
two sections. The uppermost section of length $450$ mm, width $50$ mm
and height $400$ mm functions as a hopper filled with granular
material.  The rest of the section, of length $1350$ mm, width $50$ mm
and height $200$ mm, functions as a chute. The material from the
hopper flows on to the chute through a gap of height $30$ mm which can
be opened or closed using a manually operated gate. The side walls of
the entire system are made out of stainless steel (SS$316$), except
for a transparent acrylic window of length $600$ mm and height $200$
mm, located downstream ($600$ mm from chute end) for imaging the
flow. The base of the chute (of width $50$ mm) is made from glass and
is roughened by gluing glass beads of desired diameter ($d$) on it.

Nearly spherical glass beads (Jaygo Inc., USA) of diameter $d =$ $2$
and $3$ mm, with a polydispersity of $15$\% are used as granular
material, while Magnesium Stearate (MgSt) powder (Loba Chemie, India)
is used as a lubricant material. The individual particles of MgSt
powder have plate-like structure, equivalent sphere volume diameter of
$10$ $\mu$m with polydispersity of $10$\% and are nearly frictionless
with respect to each other. For each experiment, a pre-defined, but
very tiny, quantity of MgSt powder was mixed with granular beads of
either size in a mixer. The mixer comprises an open plastic tank
(diameter $185$ mm and height $100$ mm) fitted with an overhead
stirrer. The mixing, for each experiment, was carried out by rotating
the stirrer at a speed of $30$ revolutions per minute (rpm) over a
duration of $5$ minutes. The glass beads were colored using blue ink
(Camlin Inc.) for ease in imaging and analysis. The MgSt powder, white
in color and slightly cohesive in nature, adheres itself to the
surface of the glass beads during the mixing process. The weight ratio
of MgSt to glass beads was varied between O$(10^{-5})$ and
O$(10^{-3})$ across all experiments. For a given weight ratio, the
lubricant concentration was defined $c_{l} = M_{l}/A_{p}$, where
$M_{l}$ is the total mass of lubricant, $A_{p} = n \pi d^{2}$ is the
total surface area of particles, $d$ is particle diameter and $n$ is
the number of particles. The value of $n$ is determined as
$(6/\pi d^{3}) (M_{p}/\rho_{p})$, where $M_{p}$ is total mass of
particles, $\rho_{p}$ is the particle density ($2.5 g/cm^{3}$) and
$M_{p}/\rho_{p}$ is the total particle volume.

\begin{figure}
  \centering
  \includegraphics[width=0.8\linewidth]{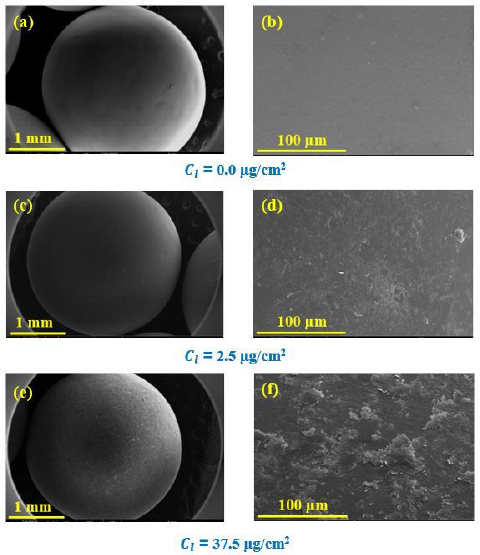}
  \caption{FE-SEM images of individual glass beads of diameter $3$ mm
    after mixing with the lubricant at different
    concentrations. Images in the right column represent the magnified
    view of the full particle images shown in the left column. The
    scales are provided for ready reference. The mixing was carried
    for $5$ minutes using a stirrer speed of $30$ rpm.}
  \label{fesem-images}
\end{figure}

Given the tiny amounts of added MgSt powder and the small
size of individual particles, it was not possible to quantify the
homogeneity of mixing as well as any loss of MgSt powder in the mixing
vessel. We, thus, resorted to visual inspection, ensuring that the
mixed material has whitish appearance and no obvious traces were left
on the mixer wall and stirrer surface. To ensure consistency of the
mixture, the mixing procedure, including mixing time and speed, was
maintained constant across all the experiments.  The surface of
particles after mixing with lubricant was imaged using FE-SEM (Field
Emission Scanning Electron Microscopy) to confirm the altered texture
and the adherence of the lubricant. Figure~\ref{fesem-images} shows
the images of $3$ mm glass beads at two different magnifications for
three different lubricant concentrations. A clean surface is visible
in fig.~\ref{fesem-images}b which gets progressively patchy with
increasing lubricant concentration as seen in
fig.~\ref{fesem-images}d and f. Multiple layers of lubricant are visible
at the highest lubricant concentration. Similar surface texture
variation is also observed for $2$ mm glass beads (not shown).

For every experiment, the chute hopper was filled upto $25$\% of its
total volume with the mixture of glass beads and MgSt while keeping
the opening closed. This ensured adequate supply of continuous feed to
achieve steady flow for reasonable duration in the experiment. The
gate was opened manually to allow gravity induced flow of material
down the chute surface. Given that our primary objective was to study
the effect of lubricant on the flow and not chute flow mechanics, we
kept all the operating parameters constant throughout, except for the
lubricant concentration being varied for each bead size. The gate was
kept in a fully opened state (height $30$ mm) across all the
experiments. The chute (glass) base was roughened with $2$ mm beads
for the flow of $2$ mm as well as $3$ mm glass beads. For a specified
chute angle, the material exiting the chute was collected directly on
a weighing scale to obtain its time dependent behavior. A small time
duration window ($4$s to $8$s) exhibited linear variation of mass
collected with time which provided the steady state mass flow rate
($m$).  The chute inclination angle was fixed at $24$ degrees and $22$
degrees, respectively, for $2$ mm and $3$ mm glass beads. A
progressive increase in the chute angles led to faster flows and
consequently reduced time duration of steady state flow. As described
next, the higher duration of steady flow rate is desirable for
acquiring sufficient number of flow images to get better statistics
for time averaged flow velocities. On the other hand, decreasing the
angles led to slower flows with increasing probability of flow
intermittency.  The chute angles specified above were chosen so as to
achieve reasonable optimality between flow duration and flow
smoothness, which does not necessarily occur at the same angle for
both particle sizes.

For the steady state flow duration, images of the flowing glass
beads were acquired simultaneously near the side wall in the downstream
region using a high speed camera (IDT Y4) (see fig.~\ref{schem}b). The
camera was positioned orthogonal to the chute wall and images were
taken at $1500$ frames per sec over a region of length $30$d and
height $20$d. A typical image of the flowing glass beads from one of
the experiment is shown in fig.~\ref{schem}c. The particle centers in
each image were obtained using the IDL (Interactive Data Language)
centroid algorithm available as open source software~\cite{idl}. The
instantaneous particle velocities in x-y plane were measured from the
particle postions in successive images and the time delay between
successive positions. Here, ``$x$-$y$'' plane represents chute wall,
``$x$'' represents flow direction and ``$y$'' represents the direction
normal to the flow. The entire flowing layer was divided into
horizontal bins of width $(w = 1$d) and length ($l = 20$d), oriented
parallel to the chute surface. The velocity ($v_{x}$ and $v_{y}$)
for each bin was obtained as an average over all particles
located within the bin and across all images. This resulted in
profiles of $x-$direction and $y-$direction velocities with distance
$y$ from the chute surface. Typically, the number of centroids
(i.e. particles) detected per bin per image was about $20$ close to
the base and decreased rapidly close to the free surface. The free
surface location was, then, defined as the bin within which not less
than $5$ particles were detected per image.  The distance of the free
surface from the chute base was defined as flowing layer thickness
($\delta$). The particle area fraction ($\phi_{a}$) in the flowing
layer was obtained as $A_{p} / A_{box}$, where
$A_{p} = n \times \pi d^{2}/4$, $A_{box} = \delta \times l$ and $n$ is
the average number of particles (or centroids) counted per
image. Within the visualisation zone itself, images were taken at
three different locations in $x-$direction, about $20$d apart from
each other. The velocity profiles for these three locations were
nearly the same suggesting of a fully developed flow. Experiments were
performed for two different particle sizes and about $8$ to $10$
different lubricant concentrations. Each experiment was repeated $4$
times to obtain the averages and to ensure repeatability of the
results.

\section{\label{results}Results \& Discussion}

\begin{figure}
  \centering
  \includegraphics[width=0.82\linewidth]{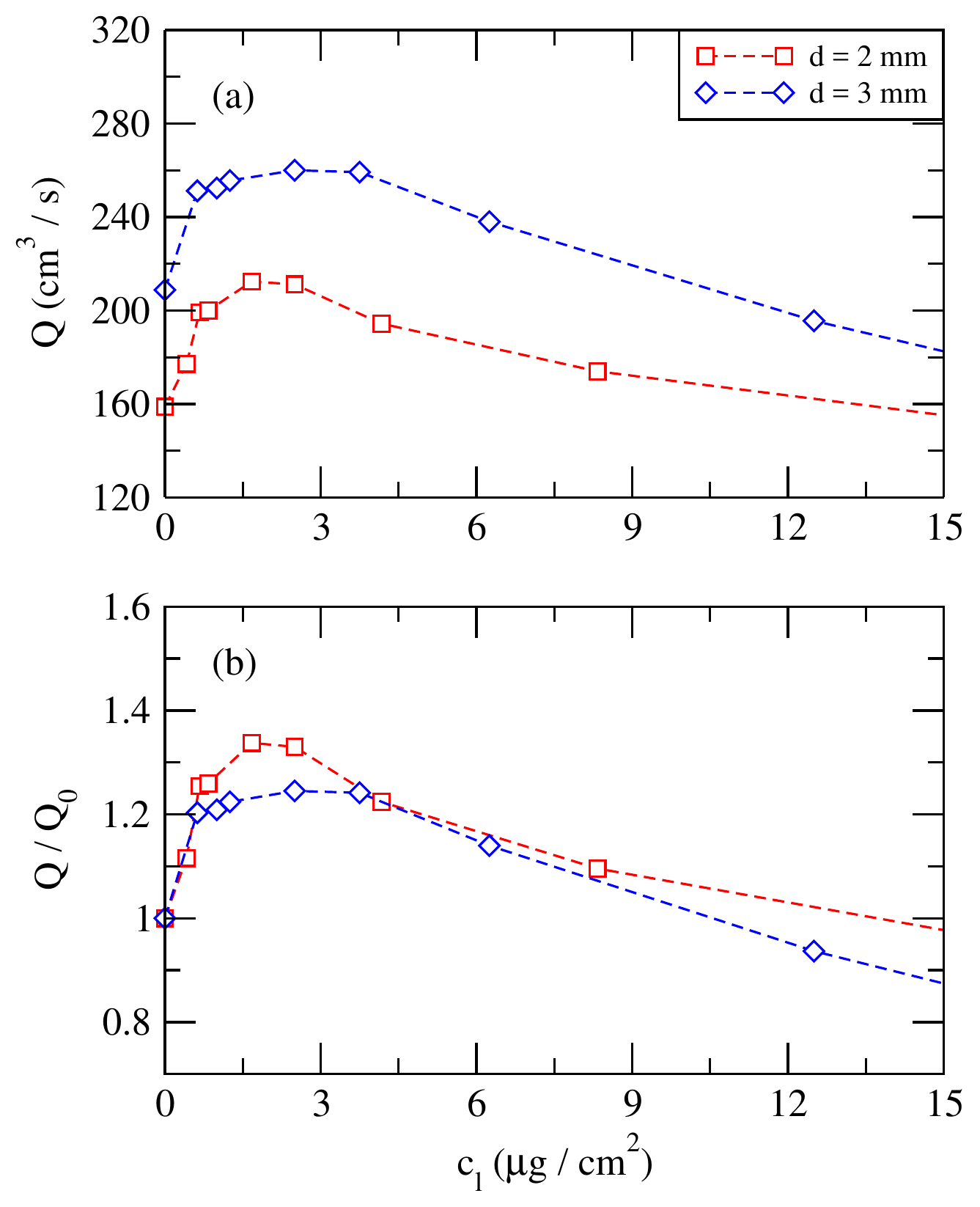}
  \caption{Variation of the volumetric flow rate with lubricant
    concentration ($c_{l}$). (a) Actual values of flow rate (Q) and
    (b) Values of flow rate (Q) normalised by its value ($Q_{0}$) in
    the absence of lubricant. Chute angle was maintained at $24$
    degrees and $22$ degrees, respectively, for flow of $2$ mm and $3$
    mm diameter glass beads.}
  \label{volflow}
\end{figure}

In the following, we discuss various attributes of flowing glass
beads on the inclined plane with respect to lubricant
concentration. The lubricant concentration ($c_{l}$) in
each case is reported as amount of lubricant material per unit surface
area of particles as described in previous section. This, then,
correctly accounts for the total area
available for the lubricant for both particle sizes considered.
The variation of the steady state volumetric flow rate ($Q$) with
lubricant concentration is shown in fig.~\ref{volflow}.  The
volumetric flow rate was obtained by dividing the mass flow rate with
particle density ($\rho_{p} = 2.5$ g$/$cm$^{3}$).  The data in
fig.~\ref{volflow}a exhibits several interesting features which we
dwell upon in the following.

Typically, the flow rate on an inclined plane, for a fixed particle
type, is uniquely determined by the inclination angle and inlet gate
opening\cite{pouliquen99}, both of which are maintained constant over
here for each particle size considered. However, a spectrum of steady
flow rate values, instead of one unique value, is observed for each
particle. This suggests of the change in the nature of the particle
surface with varying lubricant concentration. The lubricant
concentration then, possibly, acts as a parameter allowing for
studying the effect of particle surface characteristics on the flow
behavior. For a given particle size, the flow rate exhibits a
non-monotonic dependence on the lubricant concentration. It first
increases rapidly for small enough values of $c_{l}$ followed by a
slow decrease over a much wider range of $c_{l}$.  The presence of a
maximum, however, suggests that there exists an upper limit for
improving the flowability of glass beads by use of lubricant. The
particle surface characteristics which are conducive to higher
flowability at lower $c_{l}$ values seem to get altered at higher
$c_{l}$ values thereby increasing the flow hindrance instead.  For
very large enough values of $c_{l}$, the flow rate value dips below
that for base case suggesting that the presence of lubricant actually
worsens the flowability. This behavior indicates the presence of two
competing effects, the dominance of either varies at different lubricant
contents.  The maximum value of the flowrate is attained at nearly
same value of $c_{l}$ for all particle sizes, which is not quite
surprising considering the surface area driven behavior.

Figure~\ref{volflow}b shows variation of normalised volumetric flow
rate ($Q/Q_{0}$) with lubricant concentration, where $Q_{0}$ is the
flow rate in the absence of lubricant. The data shows similar behavior
and features observed in fig.~\ref{volflow}a. The profiles in
fig.~\ref{volflow}b suggests that the observed non-monotonic
dependence due to addition of lubricant can be expected to be
generally applicable for various system and operational parameters.

\begin{figure}
  \centering
  \includegraphics[width=1.0\linewidth]{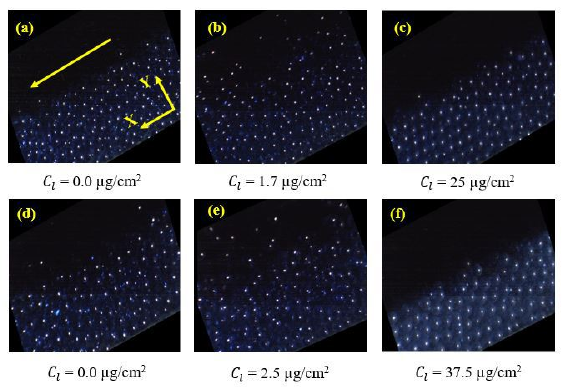}
  \caption{Images of the flowing layer of glass beads for different
    lubricant concentrations. Upper row and lower row, respectively,
    represent images for $2$ mm and $3$ mm diameter glass beads.
    Chute angle was maintained at $24$ degrees and $22$ degrees,
    respectively, for flow of $2$ mm and $3$ mm diameter glass beads.}
  \label{flow-images}
\end{figure}

Before focusing on detailed flow characteristics, we first discuss the
qualitative nature of the flow as visualised through camera
imaging. Figure~\ref{flow-images} shows images for three different
lubricant concentrations and both particle sizes. In every image
the visible base is approximately $8$ particle diameters
long. The images in the middle column correspond to the lubricant
concentration when the flow rate is maximum. The images in the left
and right column, respectively, represent the flow in the absence of
lubricant and highest lubricant concentration. Sequence of several
such images, for every particle size and $c_{l}$ values, were used to
obtain velocity data presented later on. The images clearly show a
qualitative change for varying lubricant concentration. The flowing
layer thickness increases for $c_{l} = 2.5\;\mu$g$/$cm$^{2}$ than that
in the absence of lubricant.  The flowing layer appears expanded and
seems to be in a fluidized state with particles packed thinly compared
to the base case. This leads to a larger flowing layer thickness.  The
exact opposite behavior is seen for largest value of $c_{l}$. The
layer appears to be much more densely packed, even more than that for
the base case and the thickness is reduced. Thus, the non-monotony
observed earlier for flowrate, seems to be preserved for packing
fraction as well as layer thickness. More details of the flow behavior
for all the cases shown in fig.~\ref{flow-images} can be seen in the
movies provided as Supplemental material. As evidenced from the
movies, the flow slows down substantially for the highest value of
$c_{l}$, while it is much faster for intermediate value of
$c_{l}$. The behavior then follows the classical Reynolds dilatancy
theorem which states that the material has to dilate to flow faster
(see fig.~\ref{flow-images}).

\begin{figure}
  \centering
  \includegraphics[width=0.8\linewidth]{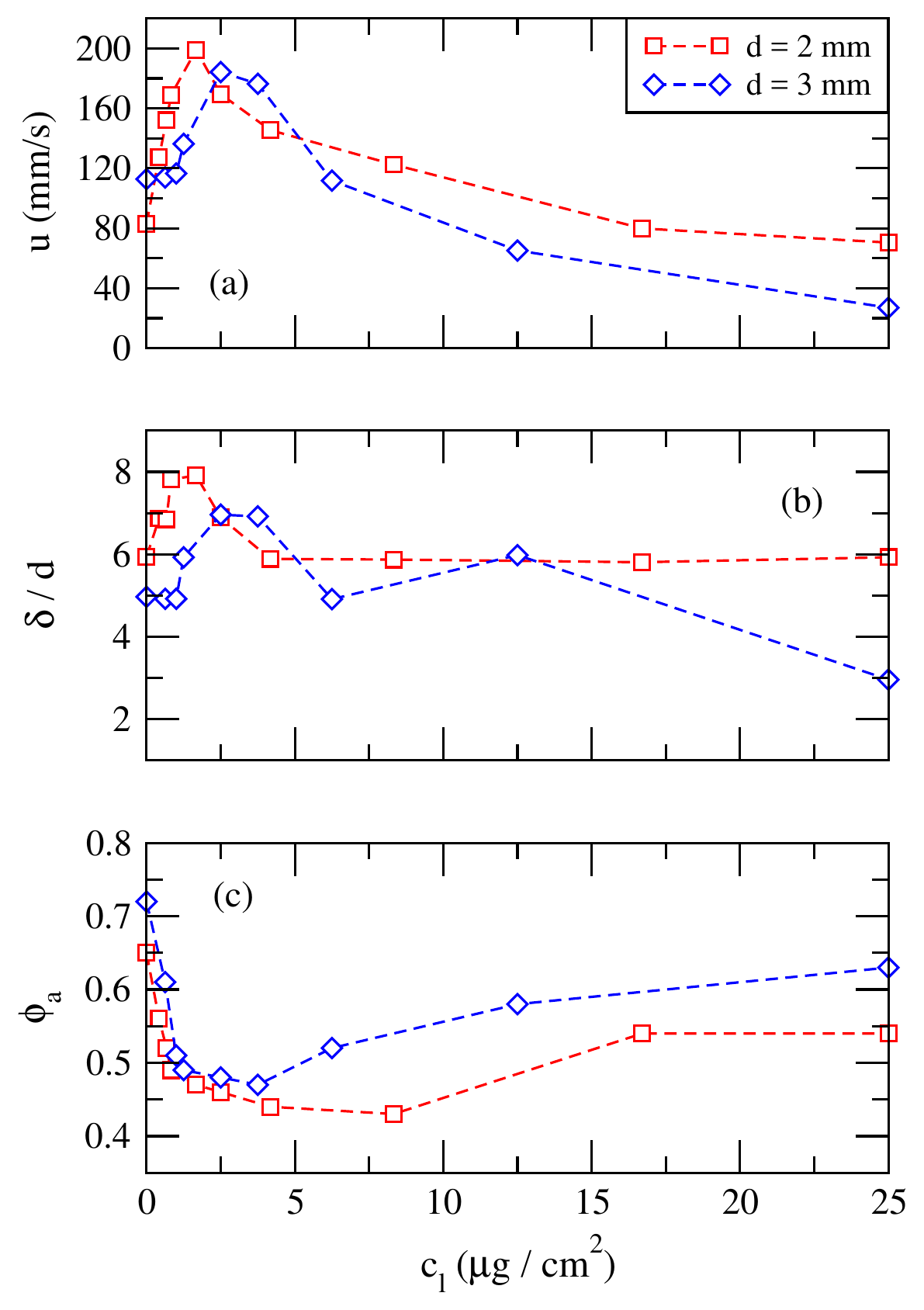}
  \caption{Variation of (a) flow depth average velocity ($u$) of
    particles in the flowing layer (b) flowing layer thickness
    ($\delta/d$) and (c) area fraction ($\phi_{a}$) measured near the
    walls with lubricant concentration ($c_{l}$). Chute angle was
    maintained at $24$ degrees and $22$ degrees, respectively, for
    flow of $2$ mm and $3$ mm diameter glass beads.}
  \label{del-vx-afr}
\end{figure}

The volumetric flow rate can be expressed as
$Q = u \delta \phi_{v} W$, where $u$ is the mean velocity in the
flowing layer, $W$ is the flowing layer width, $\delta$ is the flowing
layer thickness and $\phi_{v}$ is the mean particle volume fraction in
the flowing layer. Given that $W$ is maintained constant, the flow
rate variation will be governed by changes in the remaining three
variables for various lubricant concentrations. Since all the
measurements were carried out near the side walls, we report variation
of particle area fraction ($\phi_{a}$) in the following instead of
volume fraction ($\phi_{v}$). The flow direction velocities ($v_{x}$)
were averaged across the entire flowing layer thickness ($\delta$)
after subtracting the slip velocity ($v_{s}$) at the chute base to
obtain the depth averaged mean velocity ($u$) in the flowing
layer. The variation of the mean velocity ($u$), with lubricant
concentration is shown in fig.~\ref{del-vx-afr}a. The mean velocity
variation nearly mimics the trend for the volumetric flow rate shown
in fig.~\ref{volflow}. It increases rapidly for small enough values of
$c_{l}$ and then decreases slowly over larger values of $c_{l}$, with
a maximum at an intermediate value of $c_{l}$. On the other hand, the
layer thickness ($\delta$) for both particle sizes shows an increase
for lower values of $c_{l}$ and it almost remains constant at higher
values of lubricant concentration (see
fig.~\ref{del-vx-afr}b). Consequently, the area fraction ($\phi_{a}$)
shows rapid decrease for lower $c_{l}$ values before increasing at
higher values of lubricant concentration (see
fig.~\ref{del-vx-afr}c). These behaviors of layer thickness and area
fractions are quantitative representations of what is observed in
fig.~\ref{flow-images}. 

The lubricant (MgSt) particles are cohesive in nature and on mixing
adhere to the surface of the glass beads (see
fig.~\ref{fesem-images}). The magnified images show that the lubricant
particle coats the glass beads in patches which seem to increase in
size and thickness at very large concentrations.  Given their nearly
frictionless nature, the lubricant (MgSt) particles on adhering to the
glass beads are expected to reduce the friction coefficient between
glass beads at contact. The increase in the flow rate can, then, be
attributed to a decrease in the average inter-particle friction
coefficient, leading to a faster flow direction velocity. This leads
to enhanced collisions accompanied by lesser dissipation of particle
kinetic energy causing expansion of the flowing layer. We allude to
average value of friction over here, given that not all particles can
be coated equally by the lubricant and also not every
inter-particle contact will have lubricant presence. For large
enough concentrations, the glass beads get coated by the lubricant
particles (see figs.~\ref{fesem-images}f). While this additional
coating cannot be expected to further reduce inter-particle friction,
it can very well increase the dissipation during particle collisions,
thereby slowing down the flow while nullifying any advantage due to
reduced friction. Note that the lubricant particles are also quite
soft (malleable) in nature. The increased dissipation also leads to
flowing layer contraction. While this dissipation due to
inter-particle collision exists for all values of $c_{l}$, its effect
seems to get more pronounced with increasing lubricant
concentration. The non-monotonic behavior in that case can be
considered to arise due to competing effects of inter-particle
friction and collision, leading to a maximum at intermediate values of
$c_{l}$. The damping effect progressively increases such that at highest
concentration studied the flow rate reduces than the base case
value. With further increase in the value of $c_{l}$, the flow can be
expected to cease completely due to excessive inter-particle damping.
 
\begin{figure}
  \centering
  \includegraphics[width=1.0\linewidth]{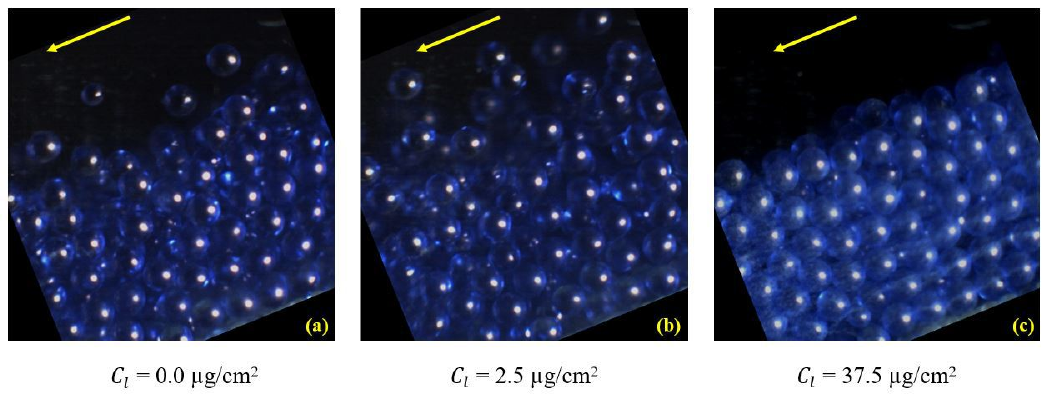}
  \caption{\textcolor{blue}{Magnified images of the flowing layer of $3$ mm
      diameter glass beads for different lubricant concentrations
      showcasing the state of particle packing. Chute angle was
      maintained at $22$ degrees.}}
  \label{flow-images1}
\end{figure}

The presence of lubricant also seems to induce microstructural changes
in the flowing layer giving rise to varying degree of compactness
which is quantified in fig.~\ref{del-vx-afr}c and visualised through
magnified images shown for $3$ mm diameter glass beads in
fig.~\ref{flow-images1}. Similar behavior is also observed for $2$ mm
glass beads (not shown). For the base case, i.e. in the absence of
lubricant, the layer shows reasonable packing particularly in the few
layers near the base. The layer seems to comprise separate clusters of few
particles throughout without connectivity across the entire layer. For
intermediate lubricant concentration (see fig.~\ref{flow-images1}b),
the flowing layer expands and the interconnectivity between the beads
is diminished even further. The flowing layer, in this case, is
dominated by inter-particle collisions.  For the highest lubricant
concentration studied, the flowing layer compacts substantially with
significant connectivity lasting over several beads instead of few
clusters as evident for lesser lubricant concentrations. Moreover, the
flowing layers shows the beads arranged in layers which move over one
another (see movie provided as supplementary material), not quite
evident at lesser lubricant concentrations. We conjecture that this
compaction and layering arises due to enhanced damping during
collisions at higher lubricant concentrations. While it cannot be
ascertained quantitatively, the layering may be associated with load
bearing stress chains spanning across the layer which may govern the
overall flow behavior.

The above alluded interactions are not feasible to be measured at
individual particle level given that the actual interactions can be
between two or more particles. Further, it may not provide reasonable
insight given the variation in the lubricant coating across several
glass beads. Since all the observations are made at the flow (bulk)
scale it is necessary to determine the quantity that can describe the
bulk behavior appropriately. This quantity will, then, represent an
effective or an average of actual interactions occurring at the
individual particle level, but manifested at the bulk level.

To quantify the effective (or average) interactions between lubricant
coated glass beads we measure the static angle of repose
($\theta_{r}$). To carry out these measurements we slowly pour mixture
of glass beads and lubricant of various concentrations in a
rectangular cell of height ($21$ cm), length ($48$ cm) and width
($3.5$ cm) to form a static heap. The length and height are
sufficiently large so as to prevent any kind of end effects. The
static image of the heap was captured using a digital camera
positioned sideways and orthogonal to the side wall of the rectangular
cell. Central, nearly flat, free surface, region of the heap (about
$20$ cm long) was analysed to obtain the angle of repose. Every experiment
was carried out $6$ times to ensure repeatability. Note that the tangent of
the angle of repose ($\tan\theta_{r}$) can be considered as the effective friction
coefficient. We conjecture that this angle should be the
result of interactions spanning several particles and is an inherent
property of the material. In essence, it should capture the countering
effects of varying inter-particle friction and inter-particle damping.

\begin{figure}
  \centering
  \includegraphics[width=0.8\linewidth]{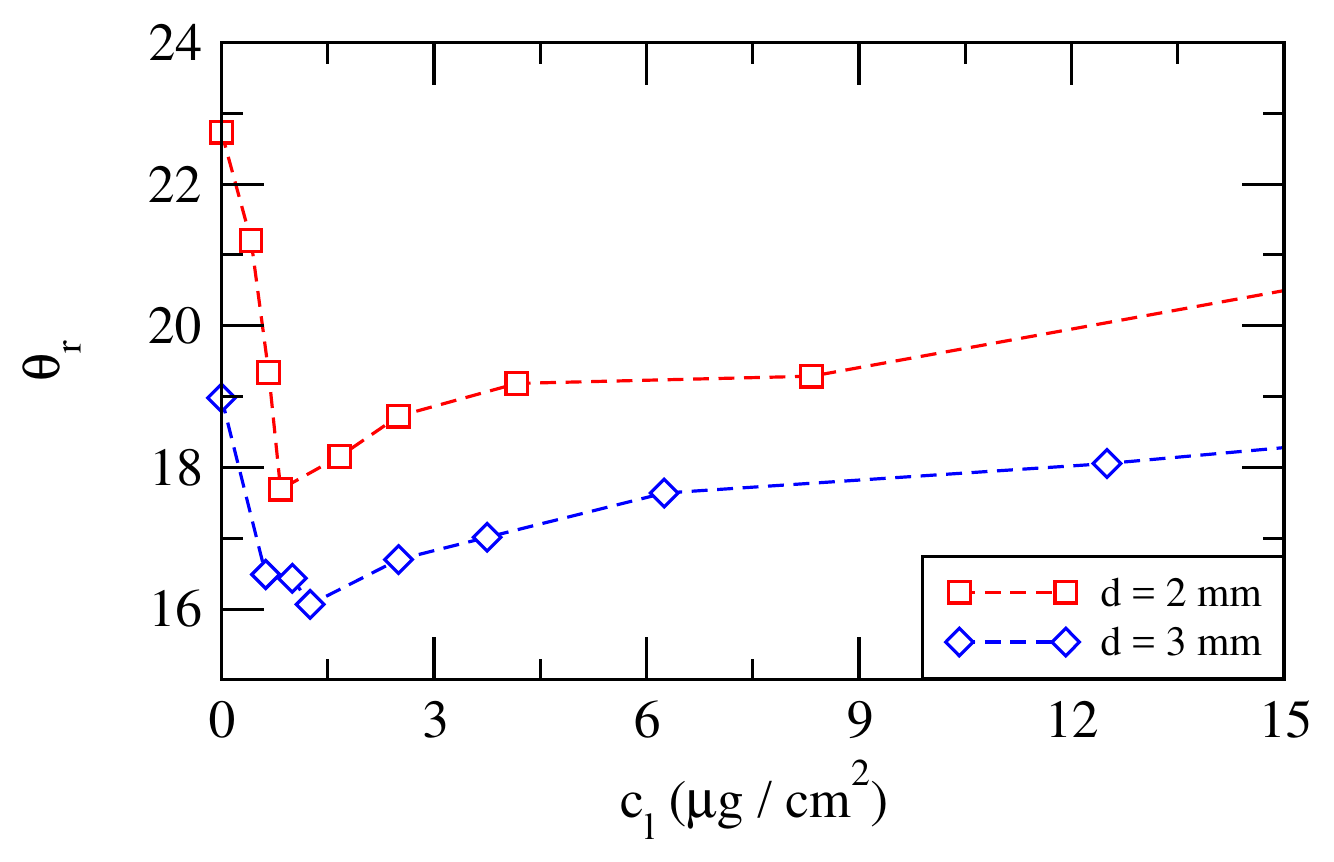}
  \caption{Variation of static angle of repose ($\theta_{r}$) with
    lubricant concentration ($c_{l}$) for the two particle sizes.}
  \label{angler}
\end{figure}

The variation of the static angle of repose ($\theta_{r}$) with
lubricant concentration is shown in fig.~\ref{angler}. The observed
behavior correctly captures the flow rate and velocity curves with
respect to lubricant concentration. The static angle of repose
(i.e. effective friction coefficient) also shows a non-monotonic
dependence on lubricant concentration ($c_{l}$) with the location of
minimum at same value of $c_{l}$ for both particle sizes. For small
enough lubricant concentrations, the value of $\theta_{r}$
progressively decreases and reaches a minimum. This indicates that the
heap of material flows relatively easily and settles at a lower angle
of repose. Correspondingly, the material progressively flows faster on
the chute surface leading to the increasing part of the flow rate (see
fig.~\ref{volflow}) and velocity curves (see fig.~\ref{del-vx-afr}a).
However, for large enough concentrations, the flowability
progressively reduces with pile settling at higher angles. This
behavior seems more likely to arise due to increased damping of
softer, highly coated glass beads rather than increased inter-particle
friction. Not surprisingly, flow rate and mean velocity curves show a
progressive decrease at larger lubricant concentrations. In fact for
large enough values of $c_{l}$, the flow rate reduces below the base
case value (in absence of lubricant). It can, thus, be expected that
the flow will completely cease to stop at very high lubricant
concentrations with the value of angle of repose (from heap measurements)
more than the inclination angle of the chute. Similar qualitative
behavior the angle of repose was also previously observed by Morin and
Briens~\cite{morin13} for various types of lubricants. The study
showed that the angle of repose decreased to a lower value for smalle
lubricant concentrations, but unlike the observations over here, it
stayed nearly at the same value for even higher lubricant
concentrations. We attribute these difference to the type of granular
particles (very small sized, cohesive pharmaceutical powders) used in
the work as against large sized, non-cohesive glass beads used over
here.

\section{\label{sum}Summary}

In summary, we have investigated the flow behavior of granular
material in the presence of a tiny amount of very small sized
Magnesium Stearate as lubricant particles. The overall behavior shows
a distinct non-monotonic dependence on increasing concentration of the
lubricant particles. For small enough lubricant concentrations, the
particles move much faster while at very large concentrations, the
flow slows down substantially. The maximum value is obtained for the
same value of lubricant per unit particle surface area for different
sized particles. The observed non-monotonic behavior is explained
through measurement of static angle of repose which represents
effective friction. We conjecture that the measured effective friction
is realised because of competing effects between (i) reduced
inter-particle friction due to frictionless lubricant adhering to the
particle surface and (ii) damping due to excessive coating of lubricant on
the particle surface thereby rendering them softer. The latter effect is
typically not observed in pharmaceutical powders, but seems to be a
signature of the hard-sphere like, non-cohesive, large sized granular
particles used over here. For such particles the normal interaction
(or collisions) do contribute substantially to the overall flow
behavior in addition to inter-particle friction. Nevertheless, the
inclined plane geometry used in this work do seem to provide an
alternative, simpler system to those typically used in pharmaceutical
powder flow characterisation. It, however, remains to be seen if the
observed behavior over here carries over smoothly to the cohesive,
pharmaceutical, small sized fine powders.

\section*{Acknowledgements}
  AVO gratefully acknowledges the financial support from Science \&
  Engineering Research Board, India (Grant No. CRG/2019/000423).

\section*{Supplementary data}
The supplementary material (movie files and their information) related
to this article are also uploaded to arXiv.

\section*{References}

\bibliography{chute}

\end{document}